\documentclass{article}

\usepackage[english]{babel}

\usepackage[letterpaper,top=2cm,bottom=2cm,left=3cm,right=3cm,marginparwidth=1.75cm]{geometry}

\usepackage{amsmath}
\usepackage{amssymb}
\usepackage{graphicx}
\usepackage[colorlinks=true, allcolors=blue]{hyperref}

\DeclareRobustCommand{\stirling}{\genfrac\{\}{0pt}{}}

\title{Exploring Network Structure with the Density of States}

\author{Rudy Arthur\\
{University of Exeter, Department of Computer Science,}\\
{Stocker Rd, Exeter EX4 4PY}\\
{E-mail:  R.Arthur@exeter.ac.uk}
}

\begin{document}
\maketitle

\begin{abstract}
Community detection, as well as the identification of other structures like core periphery and disassortative patterns, is an important topic in network analysis. While most methods seek to find the best partition of the network according to some criteria, there is a body of results that suggest that a single network can have many good but distinct partitions. In this paper we introduce the density of states as a tool for studying the space of all possible network partitions. We demonstrate how to use the well known Wang-Landau method to compute a network's density of states. We show that, even using modularity to measure quality, the density of states can still rule out spurious structure in random networks and overcome resolution limits. We demonstrate how these methods can be used to find `building blocks', groups of nodes which are consistently found together in detected communities. This suggests an approach to partitioning based on exploration of the network's structure landscape rather than optimisation. 
\end{abstract}

\section{Introduction}\label{sec:intro}

In many applications nodes in a network are partitioned into a relatively small number of non-overlapping groups. The most popular type of partitioning is to group the nodes into communities, where there are more edges within groups than between them \cite{radicchi2004defining}. There is a vast literature on methods for finding these partitions, a small sample of the better known algorithms \cite{newman2004finding,blondel2008fast,lancichinetti2011finding, rosvall2008maps} has tens of thousands of citations. It is also increasingly recognised that other types of network partition are informative \cite{liu2023nonassortative,arthur2024constraints}, the most important of these being core-periphery \cite{borgatti2000models} and disassortative structures \cite{holme2003network}. 

When grouping  nodes in a network we have to consider
\begin{enumerate}
    \item The number of groups.
    \item The structure of the groups e.g. assortative or disassortative.
    \item The quality of the label assignment.
    \item The statistical significance of the result.
\end{enumerate}
The third point has probably received the most attention in the literature, with many evaluations comparing the ability of different algorithms to find planted community structure \cite{yang2016comparative}. However, a growing understanding of the diversity of network structures as well as the need for more rigour in practical application \cite{peel2022statistical, yanchenko2024generalized} motivates us to  develop methods of graph partitioning that address all of these points. 

Our approach is based on an efficient way to sample the space of all possible networks, labellings and structures that can be generated under some null model. This method could be seen as a frequentist alternative to Bayesian methods which address the same issues \cite{peixoto2019bayesian}. These likelihood based methods assume the observed network is the output of a model - usually the degree corrected Stochastic Block Model - and seek to find the model parameters, including the node labels, which maximise the likelihood, or the posterior probability when using some prior, of the observed network under that model. In this paper, instead of trying to fit a model to the data we compare the observed network and its partitions to an ensemble of null graphs to determine if a partition, or a network, is likely or not within that ensemble.

The tool that will enable these comparisons is the density of states (DOS), which counts the number of states, in our case partitions of the network, which result in the same value of some quality function. Similar ideas were explored by other authors e.g. \cite{karrer2008robustness} to test the `robustness' of community structure. The approach described here is close to the thermodynamic method of \cite{massen2006thermodynamics} who probe the partitions of a network as a function of a temperature parameter using a Metropolis Monte Carlo approach. The key difference in our work is that focusing on the density of states allows us to use much more efficient, non-Markovian, sampling algorithms which can rapidly cover the sample space and yield a function which can characterise a network.

In Section \ref{sec:dos} we introduce the density of states, in Section \ref{sec:nulls} we introduce the relevant null models and the quality function we will use. In Section \ref{sec:WL} we discuss using the Wang-Landau algorithm to estimate the density of states. Section \ref{sec:results} applies this method to a number of illustrative examples and we discuss these findings and their implications in Section \ref{sec:outro}.

\section{Density of States}\label{sec:dos}
 For a sample space $\Omega$ with a quality function $Q : \omega \in \Omega \rightarrow \mathbb{R}$ the density of states (DOS) $g(q)$ is the number of elements $\omega$ of $\Omega$ where $Q(\omega) = q$ for some state $\omega$. The density of states is a standard tool in condensed matter physics \cite{kittel2018introduction} where $Q$ is usually the energy of the system. Formally
 \begin{align}
    g(q; \Omega) &= \sum_{\omega}^{\Omega} \delta(q, Q(\omega) ) \\
    \hat{g}(q; \Omega) &= \frac{g(q; \Omega)}{\sum_q g(q; \Omega)}
 \end{align}
where the normalised form, $\hat{g}$, is the probability that a state has quality $q$, and the sum in the denominator is over all possible values of $Q$, which is assumed to be a bounded function. In statistical physics $g$ is often used to turn multi-dimensional sums over phase space into one-dimensional sums over energy, though is also an interesting object of study in its own right.

Given a state $\omega$, a small value $\hat{g}(  Q(\omega) )$ indicates that the state $\omega$ has an unusual $Q$ value. For example, if we look at the space of all $K$-partitions of a graph, $\Omega(c; G,K)$, then $\hat{g}(q; \Omega(c; G,K)) \equiv \hat{g}_c$ tells us how likely a random partition is to have a quality $q$. One can also compare densities of states on different sample spaces, $\Omega$ and $\Omega'$, to see if there is a meaningful difference between values of $Q$ which could be sampled from those spaces. For example, we can compute $\hat{g}_{c}(q)$, the DOS over the space of possible partitions and compare that to $\hat{g}_{ cm }(q)$, the DOS over the space of all possible partitions of networks with the same degree sequence as $G$. If $\hat{g}_c(q)$ and $\hat{g}_{ cm }(q)$ are meaningfully different then this is evidence that the observed network has some structure which is probed by $Q$ and is not present in the null.

One could achieve a similar outcome by generating many random rewirings of the observed network, finding $c$ with some optimisation algorithm and comparing the distribution of $Q$s obtained with the optimal $Q$ found on the real network. However it is now recognised that the single best partition of a network does not tell the whole story, with networks often having multiple `good' but distinct partitions \cite{good2010performance,riolo2020consistency,lee2021consistency,kirkley2022representative}. Approaching questions of significance through the DOS allows us to use not just the optimal partition, but also the many sub-optimal but still informative partitions. The DOS can also be a useful summary of the network in its own right as we will see.

\section{Null Models and Quality Functions}\label{sec:nulls}

For a network $G$ let $c$ be a labelling function which assigns a number from $1$ to $K$ to each node in $G$. Our null model for partitions is that any random labelling with $K$ distinct labels is equally likely. Let $\Omega(c; G,K)$ be the space of all possible labellings of the nodes of $G$ with $K$ labels. If the network has $N$ nodes, the number of such labellings is $\stirling{N}{K}$, a Stirling number of the second kind. These numbers grow very rapidly with $N$, e.g. $\stirling{N}{2} = 2^{N-1}-1$ so a direct enumeration is impossible for any reasonably large network. 
 
 We can also have a null model for the network links. With an Erdos-Renyi null model, ${\cal ER}$, $\Omega( {\cal ER}(G) )$ is the space of all networks with the same number of nodes and edges as $G$. In practice the ER null is too general and the configuration model ${\cal CM}$ \cite{molloy1995critical} is preferable. Under this null model elements in $\Omega( {\cal CM}(G) )$ are networks with the same degree sequence as $G$. 

For our quality function we will use the generalised modularity from \cite{arthur2023discovering}
\begin{align}
    Q(c; B) &= \sum_{ab}^K Q_{ab} B_{ab} \\
    Q_{ab}  &= \sum_{i :\{c(i)=a\}, j:\{c(j)=b\} } A_{ij} - \frac{k_ik_j}{2E} = S_{ab} - \frac{T_a T_b}{2E}
\end{align}
where $c$ is the node labelling, $A$ is the adjacency matrix of $G$, $k_i$ is the degree of node $i$ and $2E$ is twice the number of edges of $G$. $S_{ab}$ and $T_a$ are the sum of edges between groups $a$ and $b$ and the sum of degrees in $a$ respectively. $B$ is a $K \times K$ `structure matrix' that specifies the mesoscopic structure of the network, for example $B$ with all ones on the diagonal and $0$ elsewhere makes $Q$ equivalent to the standard modularity which targets assortative communities. Other block patterns target other network structures, see \cite{arthur2023discovering, kojaku2018core}.

We could use other quality functions \cite{chen2015new,miyauchi2016z} but modularity is well known and this generalised form allows us to conveniently study structures, like core-periphery and bipartite networks, through the matrix $B$. A simple null model for $B$ is that any of the $2^{K^2}$ possible patterns is equally likely and $\Omega( B )$ is the space of all such block matrices. Although not all block patterns correspond to allowed network structures \cite{kojaku2018core, arthur2024constraints}, in practice we find that this null still works well.

We note that the generalised modularity shares well-known issues with regular modularity, e.g. the resolution limit \cite{fortunato2007resolution, arthur2024constraints} and a tendency to overfit \cite{ghasemian2019evaluating}. However, modularity is well known and well studied compared to some of the alternatives and generally does discriminate between `good' and `bad' partitions. The DOS can be computed for any quality function, but the simple form of modularity allows local updates to be computed quickly, which is crucial for the algorithm we will use to approximate the DOS. We avoid some of modularity's problems by fixing $K$, the number of communities. Although modularity is often used to optimise both the number and composition of communities we only use it for the latter, and note that finding the number of communities is a hard problem in general \cite{von2010clustering} and in the context of networks requires its own specialised techniques e.g. \cite{riolo2017efficient}. We are also not just looking at the modularity score of a single partition on the observed network, we will be looking at a huge ensemble of partitions and networks via the DOS and as we shall see in Section \ref{sec:results} this perspective solves many of modularity's problems.

\section{Wang Landau}\label{sec:WL}
The Wang Landau algorithm is a method to estimate the density of states \cite{wang2001efficient}. Since the values of $g$ typically become very large, we work with its logarithm, which is equal to the microcanonical entropy $S(q) = \ln g(q)$. We first divides the range of $Q$ into a number of bins. Initially we use a flat density of states $g(q) = 1 \leftrightarrow S(q) = 0$ and also keep a histogram of the number of visits to each $Q$ bin, with $H(q) = 0$ initially. A transition from state $\omega$ to $\omega'$ is proposed with probability $\pi(\omega \rightarrow \omega')$ and accepted with probability
\begin{align*}
    P(\omega \rightarrow \omega') = \min \left( 1, \frac{g( Q(\omega) )}{g( Q(\omega') )} \frac{\pi(\omega' \rightarrow \omega)}{\pi(\omega \rightarrow \omega')}\right)
\end{align*}
After this step, call the the quality of the current state $q$. The entropy in this bin is incremented by a factor $f$ and the histogram by a factor 1
\begin{align*}
    H(q) \rightarrow H(q) + 1 \quad S(q) \rightarrow S(q) + f
\end{align*}
Increasing $S(q)$ makes states with that value of $q$ less likely to be sampled and repeating the process forces the random walk to visit all allowed values of $Q$ equally often. When the histogram is sufficiently flat, the values $H(q)$ are reset to zero and the update factor $f$ is reduced, usually $f \rightarrow f/2$.

Assuming a symmetric proposal distribution, the ratio of $\pi$s in the acceptance probability cancels out. In that case we have
\begin{align*}
   P( Q(\omega) \rightarrow Q(\omega') ) = \min \left( 1, \frac{g( Q(\omega) )}{g( Q(\omega') )} \right)
\end{align*}
where we changed the argument of $P$ since it only depends on the Q value of the state. Using this we find that, when converged, the Wang-Landau algorithm satisfies the detailed balance
\begin{align*}
   \frac{ P( Q \rightarrow Q' ) }{ P( Q' \rightarrow Q ) } &= \frac{g(Q )}{g( Q' )} \\
   \frac{1}{g(Q)} P( Q \rightarrow Q' ) &= \frac{1}{g(Q')} P( Q' \rightarrow Q )
\end{align*}
We determine convergence by choosing a threshold $\epsilon$ and stopping when $f < \epsilon$. When the algorithm has converged the $\frac{1}{g(Q)}$ and $\frac{1}{g(Q')}$ terms are the equilibrium probabilities of being in states with quality $Q$ and $Q'$ respectively. The equation above shows how the algorithm spends more time in states with 'rarer' $Q$ in order to sample the spectrum uniformly.

The following implementation details are useful,
\begin{enumerate}
    \item We implement the proposal of \cite{belardinelli2007fast} to change the modification factor $f$ to $1/(t+1)$ where $t$ is the number of iterations once $f$ is smaller than $1/(t+1)$. However in practice this is moot since the condition is never encountered in our examples.
    \item The histogram flatness criteria used by \cite{wang2001efficient} is that the minimum count is at most $x\%$ less than the average, where usually $x=5$. Instead we use the criteria of \cite{zhou2005understanding}, that each bin is visited at least $N_{min} + 1/\sqrt{f}$ times where I have set $N_{min} = 10^4$.
    \item Probably the most important modification for accelerating convergence is to split the range into multiple overlapping windows \cite{vogel2013generic}. Splitting the range $[Q_{min}, Q_{max}]$ into $N_b$ bins, starting with a state with quality $Q_0$ in bin $b_0$ we run the Wang Landau algorithm in the range $[b_0 - N_{s}-N_{o}, b_0 + N_{s} + N_{o}]$.  $N_{s}$ is half the width of the window, measured in bins, and  $N_{o}$ is an `overshoot' which helps to guard against edge effects. After convergence, we discard the left and right $N_o$ bins, shift $b_0$ by $N_{step}$ and repeat until we have covered the range of $Q$ which we determine when we have no samples in the rightmost or leftmost bin. Typical values are $N_b=200$, $N_s=20$, $N_o=10$, $N_{step}=10$. 
    \item Overlapping windows are joined by finding the point at which the slopes are most similar, with the slope computed using the 5-point finite difference approximation to the derivative \cite{vogel2018practical}.
\end{enumerate}
See \cite{vogel2018practical,chevallier2020wang} for more on the practical implementation of the Wang Landau algorithm. To estimate the DOS we need to specify the proposal function and give an efficient way to compute $Q' = Q + \Delta Q$.

\subsection{Partition Sampling}

To sample the space of partitions we use a label swapping heuristic. First select a node at random, then change its label to one of the $K-1$ labels it doesn't have, unless that node is the only member of its community, in which case do nothing. The proposal probability is $\frac{1}{N_s}$ to make no change and $\frac{1}{N-N_s} \frac{1}{K-1}$ to change a label, where $N_s$ is the number of nodes in singleton communities and is symmetric. The change in $Q$ from a move where the label of node $i$ is changed from $a$ to $b$ is
\begin{align}
    \Delta Q &= \sum_{x \in c} (S_{ix} - \frac{k_i T_x}{2E}) )(B_{xb} + B_{bx}) \\ \nonumber
    &- \sum_{x \in c} (S_{ix} - \frac{k_i T_x}{2E})(B_{xa} + B_{ax}) \\ \nonumber
    &+\frac{1}{2E} \left( A_{ii} (B_{bb} - B_{aa}) + \frac{k_i^2}{2E}(B_{ab} + B_{ba} - B_{aa} - B_{bb}) \right)
\end{align}

\subsection{Configuration Model Sampling}

To sample the configuration model of the observed network $G$ we use an edge swapping heuristic. We select two random edges $(i,j) \neq (k,l)$ and swap the ends $(i,j) \rightarrow (i,l)$, $(k,l) \rightarrow (k,j)$. We allow loops and multi-edges, as is generally done with the configuration model and is valid when the degrees are small compared to the network size \cite{squartini2011analytical}. The chance of picking any pair of edges is $\frac{1}{E(E-1)}$, and the proposal probability is symmetric. There is no change in node labels or node degrees, so the change in $Q$ is 
\begin{align}
    \Delta Q = \frac{2}{2E} \left( B_{ c(i) c(l) } + B_{c(k) c(j) } - B_{ c(i) c(j) } - B_{c(k) c(l) }\right)
\end{align}

\subsection{Structure Matrix Sampling}

We specify structure using a matrix $B$ where $B_{ab} = \pm 1$ with positive signs rewarding an excess of edges between $a$ and $b$ (or within $a$ if $a=b$) and negative signs rewarding a deficit. To sample the null where all structures are equally likely we  choose a random pair of communities $a,b$ (where $a = b$ is allowed) and swap the sign of $B_{ab}$ (and $B_{ba}$ if $a \neq b$). This is a symmetric proposal step and
the change in $Q$ if block $a,b$ is flipped is
\begin{align}
  \Delta Q =  \frac{1}{2E} (2 - \delta_{ab}) \left( S_{ab} - \frac{T_a T_b}{2E} \right)
\end{align}
\\ \newline

\subsection{Sampling from Combined Nulls}

Call the space of all $K$-partitions of all networks with the same degree sequence as $G$, $\Omega(c; G,K) \times \Omega({\cal CM}(G))$ and the density of states $g_{cm}$. To traverse this space we make a swap move with probability $p_{swap}$ and a rewire move with probability $1 - p_{swap}$. Similarly, to explore all partitions and structure matrices, $\Omega(c; G,K) \times \Omega(B)$ with density of states $g_{cB}$ we make a swap move with probability $p_{swap}$ and a flip with probability $1-p_{swap}$. For $\Omega(c; G,K) \times \Omega({\cal CM}(G)) \times \Omega(B)$, all partitions and all structures for all networks with the degree sequence of $G$, with density of states $g_{cmB}$, we make a swap move with probability $p_{swap}$ a rewiring move with probability $p_{rewire}$ and a flip with probability $1 -p_{swap} - p_{rewire}$ where $p_{swap} + p_{rewire} < 1$.

\subsection{Entropic Sampling}\label{sec:entropic}
It will be useful to be able to sample states with quality greater than some value $Q_{min} = \alpha Q_{max}$, where $\alpha < 1$. To do this the Wang-Landau algorithm is run at constant $f$ \cite{lee1993new}. We first perform a large number of `warm up' moves to estimate $Q_{max}$ \cite{vogel2018practical}, compute $Q_{min}$, then every $N_{corr}$ steps check if $Q > Q_{min}$ and if so, record the state. $N_{corr}$ is chosen so that there are many accepted moves between checks. Typical values to be used is $10^4$. If a new $Q_{max}$ is found at any point we recompute $Q_{min}$ and restart the sampling. We use $\alpha = 0.99$.

\section{Results}\label{sec:results}

\subsection{Community Detection}

We first study the case of a partition into 2 assortative communities, $$
B = \begin{pmatrix}
    1 & -1 \\
    -1 & 1
\end{pmatrix}
$$
for some small, well-known networks: the karate club \cite{zachary1977information}, the dolphin social network \cite{lusseau2003bottlenose} and an Erd\H{o}s-R\'enyi random graph on 30 nodes with edge probability $0.2$. The blue points show $\hat{g}_c(Q)$ the density of states where the observed network is fixed and the labels are changed i.e. the null space $\Omega(c;G,K)$. The orange data shows $\hat{g}_{cm}(Q)$ where the labels and the edges can vary i.e. the null space $\Omega(c; G,K) \times \Omega({\cal CM}(G))$. For the latter we use $p_{swap} = 0.2$ though the output is insensitive to the exact value used. All examples use $\epsilon = 10^{-5}$.

\begin{figure}[ht]
    \centering
    \includegraphics[trim={0pt 10pt 0pt 10pt}, clip, width=\textwidth]{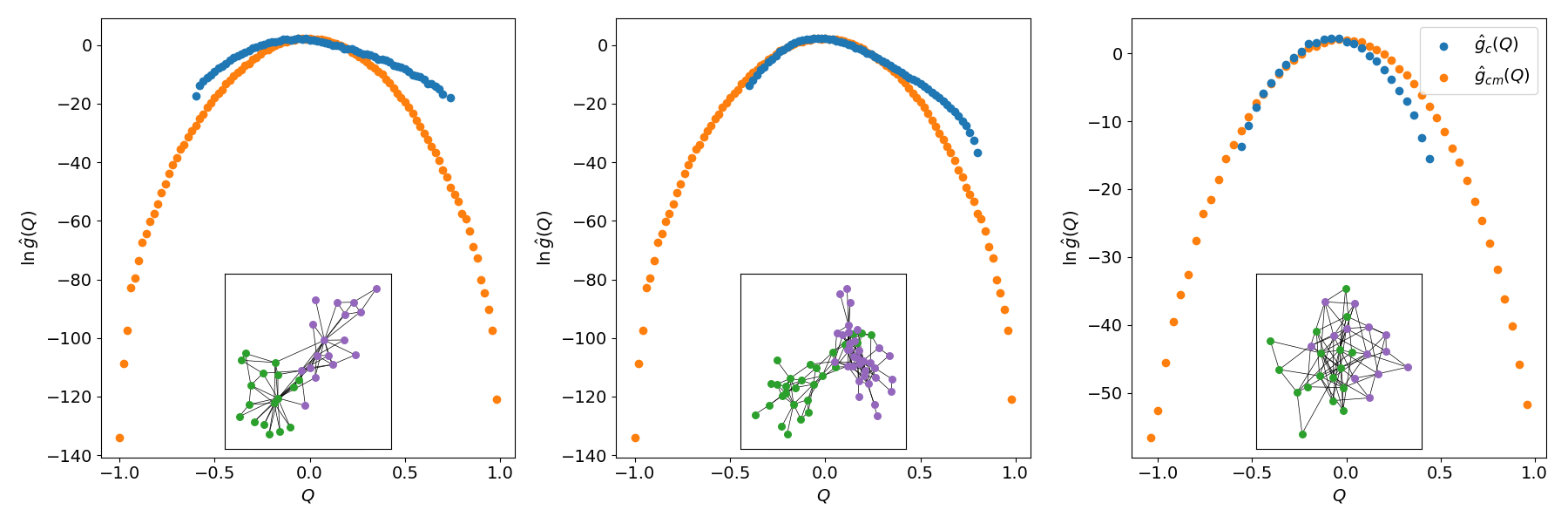}
    \caption{Left to right: karate club, dolphin social nework, ER grandom graph. Showing density of states under label swapping (blue) and label swapping configuration model (orange). Inset figure shows observed network and the `best' partition found during the Wang Landau sampling.}
    \label{fig:fig1}
\end{figure}
Figure \ref{fig:fig1} shows the corresponding densities of states. First consider the blue data - the density of states for the space of all labellings of the observed network. These all peak at $0$, the most likely $Q$ value for a random partition. The karate and dolphin networks have longer right tails, while the ER network is symmetric. The social networks have meaningful community structure (shown in the inset figure), so many states have high $Q$ values. In the ER network there are just as many high as low $Q$ states.

Looking at the orange data, the density of states in the space of all labellings and all rewirings of the observed network, these are all symmetric and cover the full range of $Q$ (the maximum value of $Q$ for a partition into $K$ parts is $2(1 - \frac{1}{K})$ \cite{arthur2023discovering}), however states with very high or very low values of $Q$ are much less likely. Note the plots show the log of the probability, each unit increment corresponds to a factor $e$ in probability. Comparing blue and orange - \emph{a high $Q$ partition is exponentially more likely in the karate and dolphin networks than in their configuration models}. We could perform some statistical test, but given the large differences in probabilities it will not be more illuminating than the qualitative comparison. Put another way, the observed network looks very different than a randomly rewired version of it in terms of community structure, with high $Q$ values much easier to find. While it is possible to find some randomisation of the network with higher $Q$, these are \emph{extremely} rare in the null space. It is very unlikely that a random network with this degree distribution has the same community structure, measured by $Q$, as the observed network. 

This is in contrast to the random graph, where the highest $Q$ partition found on the observed network is not any more likely (in fact slightly less likely) than in a randomly rewired version of it. From this we can reasonably conclude that the karate and dolphin social network have community structure and the ER network doesn't. Compare this to the well known result:  even random graphs have high modularity partitions \cite{guimera2004modularity}. Here we see that the highest modularity partition of the random network isn't statistically meaningful when compared to the configuration null. However, the highest modularity partition of the karate club or dolphin social network, both of which do have community structure, is statistically meaningful.

\subsection{Comparing Structure}

Real networks are often better described by structures other than assortative communities \cite{liu2023nonassortative}. Even the well studied karate club network has multiple `good' partitions: the usual community one (shown in Figure \ref{fig:fig1}) and an alternative `leader follower' one discussed in \cite{peel2017ground}. Specifying a target structure through the $B$ matrix we can explicitly compare competing structures. Alternatively, we can allow $B$ to vary along with the labelling and use the Wang Landau algorithm to compute the DOS, finding high quality partitions and evaluating them against the null model.

\begin{figure}[ht]
    \centering
    \includegraphics[trim={0pt 10pt 0pt 10pt}, clip, width=\textwidth]{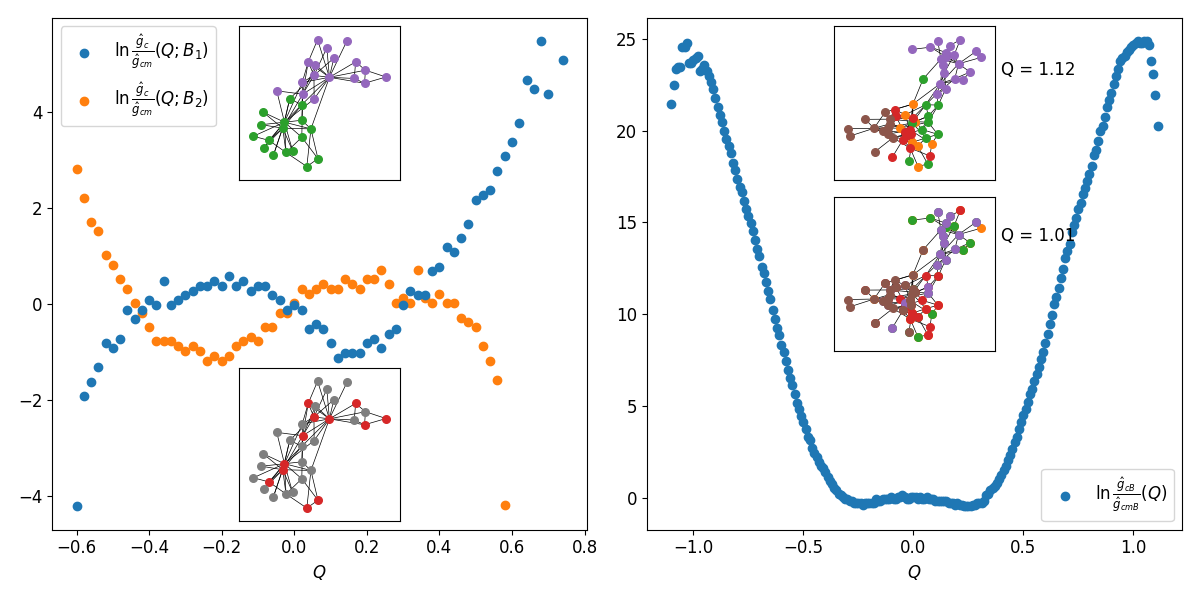}
    \caption{Left: DOS ratios for assortative and disassortative structures in the karate club network. Inset shows the highest modularity partitions for assortative (top) and disassortative case (bottom). Right: DOS ratio, where structure $B$ can also vary. Top shows the highest modularity partition (the same one found in \cite{arthur2023discovering}), centre shows a partition with highest DOS ratio. }
    \label{fig:fig2}
\end{figure}

The left panel of Figure \ref{fig:fig2} shows the ratio of DOSs where the generalised modularity is computed using 
\begin{align*}
B_1 = \begin{pmatrix}
1 & -1\\
-1 & 1
    \end{pmatrix}, \qquad \text{ and } \qquad 
B_2 = \begin{pmatrix}
-1 & 1\\
1 & -1
    \end{pmatrix}    
\end{align*}
The figure shows that high $Q$ assortative partitions, $B_1$, are much more likely than in the null, high $Q$ disassortative patterns, $B_2$, are not. The right panel shows a $K=5$ partition of the dolphins social network where the structure matrix $B$ is allowed to vary. First note there is strong support for the observed network having structure not present in the null model. The highest modularity partition is shown in Figure \ref{fig:fig2} and corresponds to the same one found in \cite{arthur2023discovering}, there are two independent communities and two communities with a shared periphery. The corresponding block pattern is
\begin{align*}
\begin{pmatrix}
. & 1 & 1 & . & . \\
1 & 1 & . & . & . \\
1 & . & 1 & . & . \\
. & . & . & 1 & . \\
. & . & . & . & 1 \\
\end{pmatrix}
\end{align*}
(where a $.$ is written in place of $-1$ to make the pattern clearer). Interestingly the highest $Q$ partitions are not the most unlikely ones, with the ratio of observed and null DOS peaking around $Q=1$. A configuration with this value of $Q$ is shown in the inset which is relatively similar to the highest $Q$ one but has structure matrix
\begin{align*}
\begin{pmatrix}
1 & . & . & 1 & . \\
. & . & 1 & . & . \\
. & 1 & 1 & . & . \\
1 & . & . & 1 & . \\
. & . & . & . & 1 \\
\end{pmatrix}
\end{align*}
with one assortative community and two different core-periphery pairs. Some aspects of the structure seem to be consistent - two densely connected assortative groups. However, among the high $Q$ partitions with $5$ groups, there is ambiguity about if the remaining three groups are in some kind of core periphery arrangement, as in the maximum $Q$ configuration, or split into three assortative communities with a small number of nodes excluded, as in the $Q=1.01$ state. We will investigate this further in the next section.

\subsection{Building Blocks}\label{sec:blocks}

Building on \cite{good2010performance} and other work on the existence of a landscape of high quality partitions \cite{riolo2020consistency} try to find the `building blocks' of these partitions. These are sets of nodes from which high quality partitions are composed. A method to identify these building blocks is given in \cite{riolo2020consistency} based on mutual information, see also \cite{kirkley2022representative} for similar ideas. 

\begin{figure}[ht]
    \centering
    \includegraphics[height=0.1\textheight, width=\textwidth]{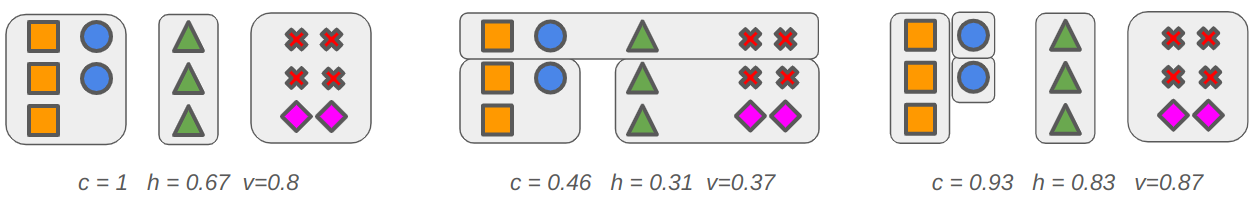}
    \caption{Building blocks (shapes) composed to create groups (grey rectangles). The left partition is perfectly complete, $c=1$, with respect to the blocks, but not homogeneous. The middle is neither homogeneous nor complete. The right partition is more homogeneous than the first, but less complete. }
    \label{fig:fig3}
\end{figure}
Here we propose to use entropic sampling to obtain high $Q$ partitions and explicitly compare using homogeneity and completeness scores of \cite{rosenberg2007v, arthur2020modularity}. If we have groups $G$ consisting of blocks $C$ the completeness is 
\begin{align}
    c = \begin{cases}
        1 \quad \text{ if } \quad H(C) = 0 \\
        1 - \frac{H(C|G)}{H(C)}
    \end{cases}
\end{align}
where $H(C|G)$ is the conditional entropy of the blocks given the groups (not to be confused with the log of the DOS). Homogeneity, $h$, is defined symmetrically with $C$ and $G$ swapped and $v = \frac{2hc}{h+c}$. If a partition is composed of building blocks, the groups of that partition should be complete with respect to the blocks, see Figure \ref{fig:fig3}. The harmonic mean of homogeneity and completeness, that \cite{rosenberg2007v} refer to as the V-measure, is closely related to mutual information,
\begin{align*}
I(C,G) &= H(C) - H(C|G) \\
c &= I(C,G)/H(C) \\
h &= I(G,C)/H(G) \\
v &= \frac{2hc}{h+c} = \frac{2I(C,G)}{H(C) + H(G)}
\end{align*}
which is often called normalised mutual information. For the purposes of finding building blocks we assume we have some large groups made of smaller blocks, as in Figure \ref{fig:fig3}. This is what is assessed by completeness. Homogeneity is less useful, for example we can make a partition more homogeneous placing a single node in its own group. Thus we prefer not to use homogeneity, or mutual information which incorporates it, instead relying on completeness to assess the quality of building blocks found.

We identify blocks with the following steps
\begin{enumerate}
    \item Generate $M=1000$ high $Q$ partitions (using $K=2$) via entropic sampling.
    \item Compute the matrix $W$, where $W_{ij}$ counts the proportion of times node $i$ is in the same group as node $j$.
    \item Find candidate blocks by placing all nodes with $W_{ij} > \theta$ into sets where $0 \leq \theta < 1$ is a threshold. 
    \item Compute the average completeness of the $M$ partitions with respect to the identified blocks as a function of $\theta$.
\end{enumerate}

\begin{figure}[!h]
    \centering
    \includegraphics[trim={0pt 10pt 0pt 10pt}, clip, width=\textwidth]{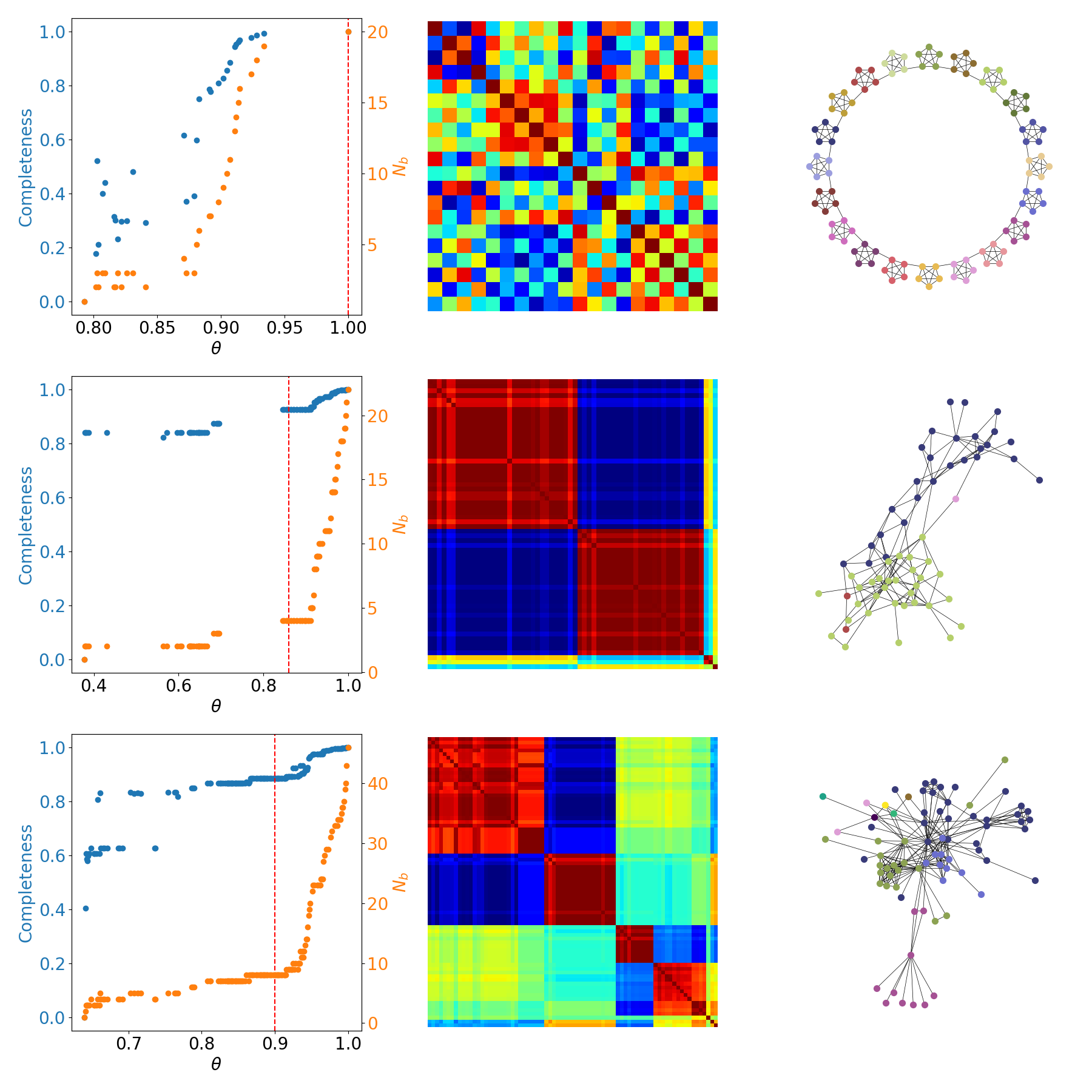}
    \caption{Left: completeness (blue) and number of blocks (orange) as a function of threshold $\theta$. Middle: the matrix $W$ sorted into blocks for the threshold indicated by the dashed red line. Right: the observed network with the building blocks coloured. Top to bottom: connected cavemen network, dolphin social network, \textit{Les Misérables} character interaction network.}
    \label{fig:fig4}
\end{figure}
In Figure \ref{fig:fig4} we do this process for the connected caveman network, where 20 cliques of 5 nodes are joined in a ring, the dolphin network and the social network of the characters in \textit{Les Misérables} \cite{knuth1993stanford}. Looking first at the caveman network - the partition with 20 blocks with completeness $1$ recovers the intuitive partition, where each clique is identified as a building block. Note this is a well known example where modularity displays a resolution limit \cite{fortunato2007resolution}, and optimising it fails to recover the cliques. Here we have found the cliques using modularity by combining a large number of optimal partitions.

The building blocks of the dolphin network for $\theta = 0.86$ are shown in the middle row of Figure \ref{fig:fig4}. This threshold indicates a partition into 4 blocks, however it is clear that there are two main blocks and a handful of nodes which are outside of them, so this network is probably best described as built from two blocks with a handful of outliers. For the \textit{Les Misérables} network we find a similar result to \cite{riolo2020consistency}, with 4 main blocks and some isolated individual nodes. 

\section{Discussion}\label{sec:outro}

In this paper we have introduced the density of states (DOS) as a tool for the analysis of community and other meso-scale structures on networks. We show that it can be used to explore various null spaces, in particular the space of all partitions, all networks with the same degree sequence and all meso-scale structures. Comparing the DOS of partitions on the observed network to the DOS of partitions of the configuration model of that network allows us to assess statistical significance. We can answer the question as to whether the high quality partitions found are unusual, giving us confidence that there is structure in the network and, by exploring the null space of network structures, what that structure is. By sampling many high-quality partitions we can identify groups of nodes which are consistently together - the building blocks of \cite{riolo2020consistency}. These building blocks are identified by a simple thresholding process and evaluated using the completeness score (conditional entropy) of \cite{rosenberg2007v}.

In this work generalised modularity is used as the quality function. When using a diagonal structure matrix $B$, this is equivalent to standard modularity \cite{newman2004finding}. Here we see that its well known issues like finding high modularity partitions in random networks \cite{guimera2004modularity} or the resolution limit \cite{fortunato2007resolution}, can be overcome with this approach. In particular, spurious communities found in random networks are not statistically meaningful, Figure \ref{fig:fig1}, and the cliques which would be merged by straightforward modularity optimisation are recovered using the approach of Section \ref{sec:blocks}.

Regarding the latter result, the building block analysis is effective, but I have deliberately avoided framing it as an optimisation problem, as in \cite{riolo2020consistency}. The results of this work, and many other precedents, suggest that straightforward optimisation algorithms don't give us the whole picture about network structure, and we don't want to replace one optimisation problem with another. Varying the threshold for placing nodes together in a block generally results in an `elbow', an abrupt change in curvature of the completeness score. Exploring the building blocks to the left of this elbow provides interesting insight into the network's structure.

For the practical application of these techniques, the entropic sampling approach is extremely fast, however while the Wang Landau algorithm is quite efficient at sampling the range of possible $Q$ values, the convergence depends on the details of the $Q$ discretisation - how many bins to use and how wide to make the windows. The algorithm can also get `stuck' at the extreme right and left of the range of $Q$, when there are bins where $g(Q)$ differs by many orders of magnitude. While it is quite fast on the small networks we study here, applying it to very large networks requires further development. The replica exchange method of \cite{vogel2013generic} or alternative methods like \cite{langfeld2012density} should be explored. The other obvious line of improvement is using quality functions other than modularity, for example variants such as \cite{chen2015new,miyauchi2016z} or likelihoods as in \cite{riolo2020consistency}. Again we leave this for future work.

This paper presents the density of states as a useful tool for understanding network structure, shows several applications of it and in particular how it solves many well known issues with modularity optimisation. However the density of states, and the algorithm for calculating it, are interesting beyond modularity and the debates over its merits. In practice, real network data may be imperfectly sampled or change over time, so the one `best' partition, measured with any quality function, may not be as useful as understanding the consistency and overlap between the huge number of `good' partitions \cite{kirkley2022representative}. This paper describes a method to explore the whole landscape of network structures and partitions, which I hope can lead to a better understanding of real world data.

\bibliographystyle{plain}
\bibliography{main}

\end{document}